\documentclass[structabstract]{aa} 

\usepackage{natbib}
\bibpunct{(}{)}{;}{a}{}{,} 
\usepackage{graphicx}
\usepackage{txfonts}

\newcommand{\zpowerlaw}{{\fontfamily{ptm}\fontseries{m}\fontshape{sc}\selectfont{zpowerlw}}}

\newcommand{\tbabs}{{\fontfamily{ptm}\fontseries{m}\fontshape{sc}\selectfont{tbabs}}}
\newcommand{\ztbabs}{{\fontfamily{ptm}\fontseries{m}\fontshape{sc}\selectfont{ztbabs}}}
\newcommand{\pexmon}{{\fontfamily{ptm}\fontseries{m}\fontshape{sc}\selectfont{pexmon}}}
\newcommand{\pexrav}{{\fontfamily{ptm}\fontseries{m}\fontshape{sc}\selectfont{pexrav}}}

\newcommand{\ky}{{\fontfamily{ptm}\fontseries{m}\fontshape{sc}\selectfont{ky}}}

\newcommand{\ekisemem}{\textit{{XMM}}}
\newcommand{\xmm}{\textit{{XMM-Newton}}}
\newcommand{\integral}{\textit{Integral}}
\newcommand{\swif}{\textit{Swift}}
\newcommand{\swift}{\textit{Swift-BAT}}
\newcommand{\suzaku}{\textit{Suzaku}}
\newcommand{\asca}{\textit{ASCA}}
\newcommand{\uhuru}{\textit{Uhuru}}
\newcommand{\pn}{\textit{EPIC-pn}}
\newcommand{\mosj}{\textit{EPIC-mos1}}
\newcommand{\mosd}{\textit{EPIC-mos2}}
\newcommand{\xis}{\textit{XIS}}
\newcommand{\hxd}{\textit{HXD/PIN}}
\newcommand{\uu}{{4U 1344-60}}
\newcommand{\cenb}{{Centaurus B}}
\newcommand{\chired}{$\chi^2_\nu$}

\begin{document}
\title{Active galaxy 4U\,1344-60: did the relativistic line disappear?}
\author{J.~Svoboda\inst{\ref{inst1}}\thanks{email: jsvoboda@sciops.esa.int}\and S.~Bianchi\inst{\ref{inst2}}\and M.~Guainazzi\inst{\ref{inst1}}\and G.~Matt\inst{\ref{inst2}}\and E.~Piconcelli\inst{\ref{inst1},\ref{inst3}}\and V.~Karas\inst{\ref{inst4}}\and M. Dov\v{c}iak\inst{\ref{inst4}}}
\institute{European Space Astronomy Centre of ESA, PO Box 78, Villanueva de la Ca\~{n}ada, 28691 Madrid, Spain\label{inst1}
\and Universit\`{a} degli Studi Roma Tre, via dellaVasca Navale 84, 00146 Roma, Italy\label{inst2}
\and Osservatorio Astronomico di Roma (INAF), via Frascati 33, I-00040 Monteporzio Catone (Roma), Italy\label{inst3}
\and Astronomical Institute, Academy of Sciences, Bo\v{c}n\'{\i}~II~1401, CZ-14131~Prague, Czech~Republic\label{inst4}}


  \abstract
   {X-ray bright active galactic nuclei represent a unique astrophysical laboratory
for studying accretion physics around super-massive black holes.}
   {4U 1344-60 is a bright Seyfert galaxy which revealed relativistic reflection features
in the archival {\xmm} observation.}
   {We present the spectroscopic results of new data obtained 
with the {\suzaku} satellite 
and compare them with the previous {\xmm} observation.
}
   {The X-ray continuum of 4U 1344-60 can be well described
by a power-law component with the photon index $\approx 1.7$ 
modified by a fully and a partially covering local absorbers. 
We measured a substantial decrease of 
the fraction of the partially absorbed radiation 
from around $45\%$ in the {\xmm} observation
to less than $10\%$ in the {\suzaku} observation
while the power-law slope remains constant within uncertainties.
%
The iron line in the {\suzaku} spectrum is relatively narrow, 
$\sigma=(0.08 \pm 0.02)$ keV,
without any suggestion for relativistic broadening. 
Regarding this, we interpret the iron line  
in the archival {\xmm} spectrum as 
a~narrow line of the same width 
plus an additional red-shifted emission 
around 6.1\,keV.
%
}
{
No evidence of the relativistic reflection
is present in the {\suzaku} spectra.
The detected red-shifted iron line during the {\xmm} observation
could be a temporary feature either due to locally enhanced emission
or decreased ionisation in the innermost accretion flow.
%
%
}



\maketitle

\section{Introduction}


Bright active galaxies provide a good
opportunity to probe accretion physics in the strong gravity regime
and to reveal the basic parameters of the accreting black hole
\citep[see e.g. ][]{2003PhR...377..389R}. 
In particular, the black hole angular momentum can be constrained 
from the relativistic features in the spectra originating
in the closest neighbourhood to the black hole.
Suitable tools for this purpose
are the high-throughput X-ray detectors on space missions
such as {\xmm} \citep{2001A&A...365L...1J} 
and {\suzaku} \citep{2007PASJ...59S...1M}
whose data analysis we are presenting in this paper.

A sizeable sample of active galactic nuclei (AGNs) 
with relativistically broadened iron lines was studied
by \citet{2006AN....327.1032G} and later by \citet{2010A&A...524A..50D}.
Different samples of Seyfert galaxies observed by the {\xmm} satellite
were also examined by \citet{2007MNRAS.382..194N}, 
\citet{2009ApJ...702.1367B}, 
and \citet{2011MNRAS.416..629B}. 
Broad iron lines in quasars were studied by \citet{2005A&A...435..449J}.
All groups concluded that X-ray spectra of a substantial fraction of AGNs 
possess a relativistically broadened iron line. The fraction increases
with longer-exposed observations providing better statistics
for spectroscopic studies.


\uu, the subject of the study discussed in this paper, 
is a bright (about 2 millicrabs) and nearby source
but it has not been intensively studied so far due to 
its low Galactic latitude ($|b|$=1.5$^{\circ}$).
Although its X-ray spectrum is heavily absorbed by the Galactic interstellar matter
(N$_{\rm H} \approx 1 \times  10^{22}$~cm$^{-2}$)
it has clearly revealed an emission excess in the broadened red wing 
of the iron line in a previous \xmm\ observation \citep{2006A&A...453..839P}. 

4U 1344-60 was discovered for the first time by the X-ray satellite
{\uhuru}.  In the \xmm\ observation of {\cenb},
4U 1344-60 appeared at the edge of the field (Obs.ID:\,0092140101). 
\citet{2006A&A...453..839P} 
studied  the {\pn} spectrum together with the optical observations of the source. 
The optical spectrum allowed for a
classification of the source as an active galaxy
of an intermediate type. The cosmological redshift 
was derived to be $z=0.012\pm0.002$.
Its flux, $f^{XMM}_{2-10\,{\rm keV}}=3.6\times10^{-11}$~erg\,cm$^{-2}$\,s$^{-1}$, 
makes it one of the brightest AGN in the hard X-ray sky.
The luminosity,  $L^{XMM}_{2-10\,{\rm keV}}=1.5\times10^{43}$~erg/s, 
is typical for Seyfert galaxies.
{\uu} was observed by  {\integral}
\citep{2010ApJS..186....1B} with its hard X-ray flux 
$f^{\rm Integral}_{20-40\,{\rm keV}}=(3.18 \pm 0.08) \times10^{-11}$~erg\,cm$^{-2}$\,s$^{-1}$
and $f^{\rm Integral}_{40-100\,{\rm keV}}=(4.1 \pm 0.2) \times10^{-11}$~erg\,cm$^{-2}$\,s$^{-1}$. 
It was also detected in the {\swift} survey
with 
$f^{\rm Swift/BAT}_{15-150\,{\rm keV}} \approx (9.0 \pm 0.3) \times10^{-11}$~erg\,cm$^{-2}$\,s$^{-1}$
\citep{2010A&A...524A..64C}.

Residuals in the energy range, where emission features
by iron line are expected, were formerly analysed 
by \citet{2006A&A...453..839P} in the \xmm\ observation.
A model consisting of the power-law continuum
plus a narrow unresolved Gaussian line
gave an unacceptable fit.
If the line profile was allowed to be distorted
by relativistic effects the fit was significantly improved.
The inner disc radius was found to be within 10
gravitational radii ($r_{\rm g} \equiv \frac{GM}{c^2}$).
The outer one was larger  than $90\,r_{\rm g}$
with a fixed radial emissivity index $q=2.5$.
The inclination of the disk was
found to be around  $20$ degrees. 
The equivalent width of the line was about $400$ eV.

{\uu} also exhibited a complex absorption in the X-ray spectrum.
\citet{2006A&A...453..839P} suggested the presence
of a partially covering absorber to explain
a remarkably flat power law describing the data.
A combined analysis of the {\xmm} and {\integral} observation
was later done by \citet{2008A&A...483..151P}
who concluded that the spectrum is affected
by one fully and two partially covering absorbers
with substantial column densities 
($N^{\rm pc1}_{\rm H} \approx 5 \times 10^{22}$ cm$^{-2}$,
and $N^{\rm pc2}_{\rm H} \approx 4 \times 10^{23}$ cm$^{-2}$) 
and covering fractions around $50\%$.

In this paper, we present new spectroscopic results
of the recent observation of 4U 1344-60 obtained with the {\suzaku}
satellite. We compare them with the
archival {\xmm} data, which we re-analysed. 
The paper is organised as follows.
Sect.~\ref{data} describes the data reduction
of the {\suzaku} {\xis} and {\hxd},
and also the re-analysis of the {\xmm} data.
The basic timing and spectral properties of the {\suzaku} observation 
are presented in Sect.~\ref{4U}.
We compare the spectral results from the Suzaku observation 
with the archival {\xmm} data in Sect.~\ref{xmm}.
The achieved results are discussed in Sect.~\ref{discussion},
and the main conclusions are drawn in Sect.~\ref{conclusions}. 
The timing and spectral analysis of {\cenb} is presented
in the Appendix\,A, as well as the estimate of 
its contribution to the {\uu} {\hxd} spectrum. 

\begin{figure}
 \includegraphics[width=0.5\textwidth]{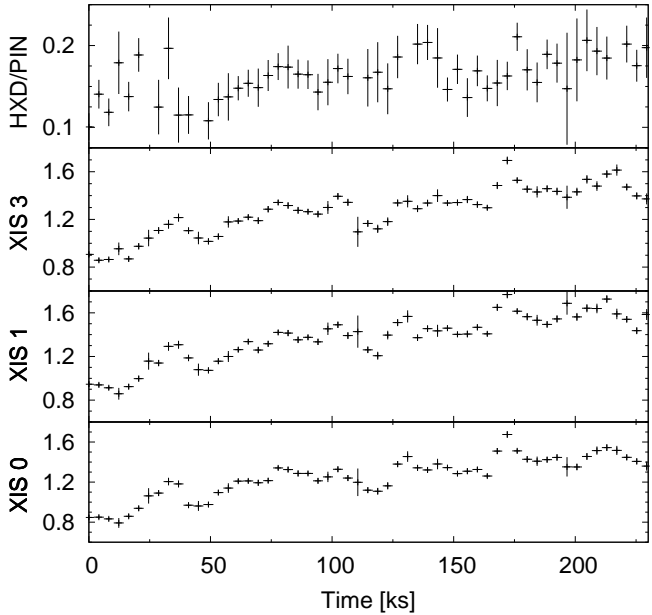}
 \caption{{\xis}\,0, {\xis}\,1, {\xis}\,3, and {\hxd} 
(all background subtracted) light curves
in the $0.5-10$\,keV, or $15-60$\,keV energy range,
respectively. The time bin size is 4096\,s.}
\label{lc_xis013}	
\end{figure}

\begin{figure}
 \includegraphics[width=0.5\textwidth]{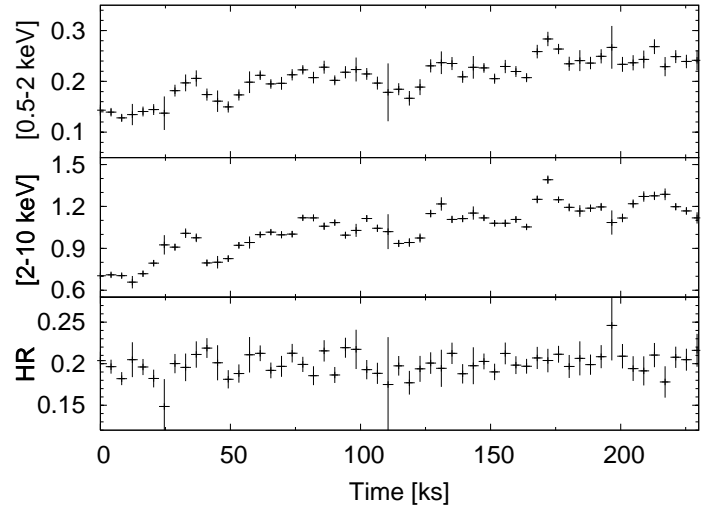}
 \caption{{\xis}\,0 (background subtracted) light curves in different spectral bands:
$0.5-2$\,keV (upper panel), and $2-10$\,keV (middle panel).
Their ratio is plotted in the bottom panel. The time bin size is 4096\,s.}
\label{lc_xis0}	
\end{figure}

\section{Data reduction and analysis}
\label{data}

4U 1344-60 was observed by the {\suzaku} satellite on January 2011 
(obs. ID 705058010) for the total exposure of 100\,ks.
The observation was performed using {\xis}-nominal position.
It was accompanied by a short 10\,ks
observation pointed to the radio-galaxy Centaurus\,B
(obs. ID 705059010), which is a neighbouring
source on the X-ray sky.
Due to its proximity, the {\hxd} spectrum of 4U 1344-60
is likely to be contaminated. 
The short observation of Centaurus\,B was performed
to measure its flux and spectral properties
below 10\,keV with the {\xis} detectors,
in order to estimate its contribution to the {\hxd}
spectrum of \uu. 
This observation strategy was possible only thanks to
the very low short-time variability of the {\cenb} 
and its simple power-law shaped X-ray spectrum, as was suggested
from the previous observations by the ASCA and {\xmm}
satellites (see Appendix). 

We used the Heasoft package version 6.11.1\footnote{http://heasarc.nasa.gov/lheasoft/}
for the data reduction and also for the subsequent spectral and timing analysis.
The data were processed standardly following the {\suzaku} Data
Reduction Guide\footnote{http://heasarc.nasa.gov/docs/suzaku/analysis/abc/} (version 4).
For all {\xis} detectors, we combined both 3x3 and 5x5 modes
to extract the event files.
The source spectra of 4U 1344-60 and \cenb\ were obtained from a circle 
around the centre of the point spread function
with the radius of 260 arcsec.
We defined the background extraction region 
as an annulus around the source circle
with the outer radius of 360 arcsec
to avoid any contamination from the calibration source.
We created the related response matrices and ancillary response files
using the tools {\textit{xisrmfgen}} and {\textit{xissimarfgen}}.
The {\hxd} spectra were reduced with the tool 
{\textit{hxdpinxbpi}}. The tuned background files
were used for modelling the non X-ray background. The standard method
was used to account for cosmic X-ray background as described in the {\suzaku} Data
Reduction Guide.


Cross-normalisation between the {\suzaku} detectors 
were fixed to 1 for {\xis}\,0, free for {\xis}\,1 and {\xis}\,3,
and fixed to 1.16 for {\hxd}. 
Fits were performed in the 1-10\,keV and 15-60\,keV energy ranges
for the {\xis} and {\hxd}, respectively. 
The data below 1\,keV were not included due to
large Galactic absorption.
The {\xis}\,1 data in the energy range 1.6-2.3\,keV were ignored
due to calibration uncertainties.
We used C-statistics \citep{1979ApJ...228..939C} 
for fitting the (unbinned) data,
which employs the appropriate Poisson distribution of count fluctuations.
However, we express the goodness of the fit also
with the more familiar $\chi^{2}$ statistics. 
Only for this purpose, 
we binned the spectra to contain at least 30 counts per bin
because the Gaussian smooth distribution of count fluctuations 
is good only for a sufficient number of counts per bin
\citep[see e.g.][for comparison of C- and $\chi^{2}$- statistics]{1989ApJ...342.1207N}.
The quoted errors in the text correspond to a $90\%$ confidence level
for one interesting parameter.


Because our spectral results with the Suzaku observation 
differ from that reported by \citet{2006A&A...453..839P} 
we also re-analysed the archival {\xmm} data
and performed a simultaneous fit to both data sets. 
We used the recent version
of the SAS software (11.0.0)\footnote{http://xmm.esa.int/sas/}.
{\uu} was observed with {\xmm} in August 2001 (obs. ID 0092140101) 
for the total exposure of 37\,ks.
The observation was aimed to Centaurus\,B.
4U 1344-60 is well detected in the {\pn} field-of-view
while it is on the edge of the {\mosj} and outside of the {\mosd} field-of-view
due to a relatively large offset ($\approx 14$\,arcmin).
Therefore, we used only the {\pn} spectra. 

The data were standardly
screened -- only patterns 0-4 were used, high background flaring
events were ignored (background rate $> 0.4$ cts\,s$^{-1}$).
The source was defined as an ellipse
(in the sky xy-coordinates: 18800.5, 12880.5, 1946.4, 909.7, 327),
which better corresponds to the shape
of the point spread function due to its offset position \citep{2011A&A...534A..34R}.
The background extraction region was defined from a nearby circle
in the source-free region of the chip.
The results of our re-analysis of the {\xmm} {\pn} 
spectrum are consistent with the results by \citet{2006A&A...453..839P}.

\begin{figure}
 \includegraphics[width=0.49\textwidth]{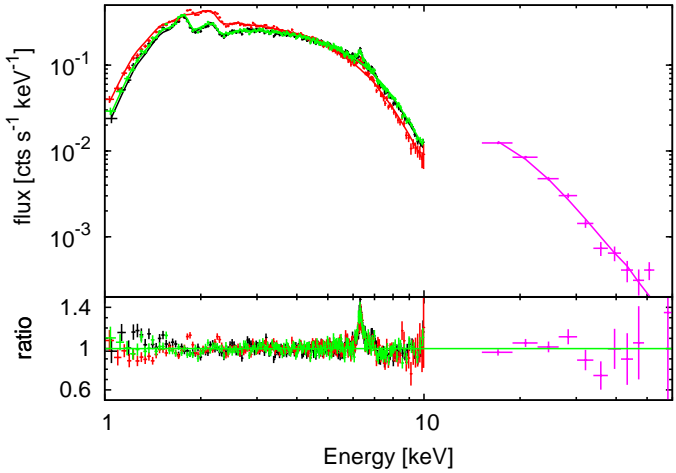}
 \caption{{\xis}\,0 (black), {\xis}\,1 (red), {\xis}\,3 (green), 
and {\hxd} (magenta) spectrum. The underlying model is an absorbed
power law with $\Gamma \approx 1.7$. 
The spectrum is binned for plotting purposes only.}
\label{tbpo}	
\end{figure}

\section{{\suzaku} view of 4U\,1344-60}
\label{4U}


The time-averaged flux of 4U 1344-60 observed by {\suzaku} is similar
to the flux measured by the {\xmm} satellite,
i.e. $f_{2-10\,{\rm keV}} \approx 4\times10^{-11}$~erg\,cm$^{-2}$\,s$^{-1}$,
or the luminosity $L_{2-10\,{\rm keV}} \approx 1.2\times10^{43}$~erg\,s$^{-1}$, respectively.
During the observation the brightness of the source increased 
by a factor of 2 - see Figure~\ref{lc_xis013} where background 
subtracted light curves of all {\suzaku} detectors are shown.
However, the spectral shape did not change significantly, 
as it is suggested by the almost constant hardness ratio 
defined as a ratio of $0.5-2$\,keV to $2-10$\,keV spectra
(see Figure~\ref{lc_xis0}).



The time averaged spectra of {\xis}\,0, {\xis}\,1, {\xis}\,3, and {\hxd} detectors
are shown in Figure~\ref{tbpo}. 
The {\hxd} spectrum of {\uu} is background-dominated
with the 27\% of the flux due to the source.
This value is well above the background systematic error 
(around 3-5\%, see e.g. Suzaku-Memo-2008-03).
The contamination of the {\uu} {\hxd} spectrum by {\cenb}
is negligible (see Sect.~\ref{hxd} in Appendix~A). 
The 15-60\,keV flux of 4U 1344-60 is approximately 
$8\times10^{-11}$~erg\,cm$^{-2}$\,s$^{-1}$ while
the estimated flux from the {\xis} spectra of Cen\,B is only $8\times10^{-12}$~erg\,cm$^{-2}$\,s$^{-1}$,
i.e. one order of magnitude lower.

The underlying model in Figure~\ref{tbpo} is an absorbed power-law
with the photon index $\Gamma \approx 1.7$. 
The {\sc{tbabs}} model \citep{wilms00} was used 
for the photoelectric absorption. 
We fixed the column density of the absorption
due to inter-stellar material in our Galaxy to the value 
$N_{H} = 1.06 \times 10^{22}$cm$^{-2}$ \citep{2005A&A...440..775K}.
The survey by \citet{1990ARA&A..28..215D} gives a consistent
value ($N_{H} = 1.05 \times 10^{22}$cm$^{-2}$).
An additional absorption of comparable column density 
was required.
We associated it to the host galaxy, although, 
regarding the very low Galactic latitude of {\uu},
some fraction of it might be due to local variations in our Galaxy.
The column density of the additional absorber was found 
to be $N_{H} = 0.93 \pm 0.02 \times 10^{22}$cm$^{-2}$ 
with the {\ztbabs} model.

\begin{figure}
 \includegraphics[width=0.49\textwidth]{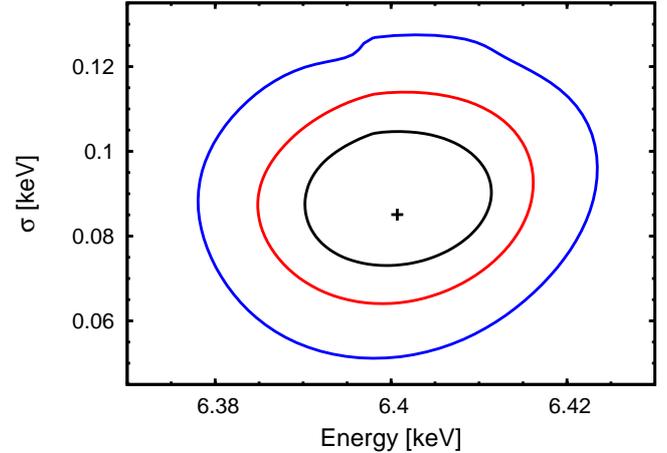}
 \caption{Contour plot of the rest energy of the iron K$\alpha$
line and its width as measured with {\suzaku}. The individual lines 
correspond to 68, 90, and 99\% confidence levels.
The best-fit values are indicated by a small cross
inside the contours.}
\label{cont_es}	
\end{figure}

The only clear residuals from this simple model are around 6.4\,keV,
and can be attributed to the iron fluorescent line. 
We added a Gaussian line to the model with the 
centroid energy and width as free parameters.
The fit is then characterised by the C-value $7615$ with $7312$ degrees of freedom ($\nu$),
or {\chired}$=5612/5452 \approx 1.03$, respectively. 
The equivalent width of the line is $96^{+14}_{-8}$\,eV.
The energy of the line was found to be $E = 6.40 \pm 0.03$\,keV 
and its width $\sigma = 0.085^{+0.025}_{-0.015}$\,keV 
(see Figure~\ref{cont_es}).

The corresponding FWHM velocity is $ \approx 8800$\,km\,s$^{-1}$,
which suggests an origin in the outermost regions 
of an accretion disc or in a broad line region
\citep{2006LNP...693...77P}.
The optical measurements of a broad component of H$_{\alpha}$
by \citet{2006A&A...453..839P} gives FWHM velocity 
twice smaller ($\approx 4400$\,km\,s$^{-1}$)
implying that the iron line must originate closer
to the black hole than the optical broad line region.
However, compared to the {\xmm} observation 
with $E \approx 6.2$\,keV, and
$\sigma \approx 0.3$\,keV \citep{2006A&A...453..839P}
the {\suzaku} iron line profile is much narrower. 
Therefore, we perform a simultaneous fit
of both observations in the next section.




\begin{figure*}
 \centering
\includegraphics[width=0.9\textwidth]{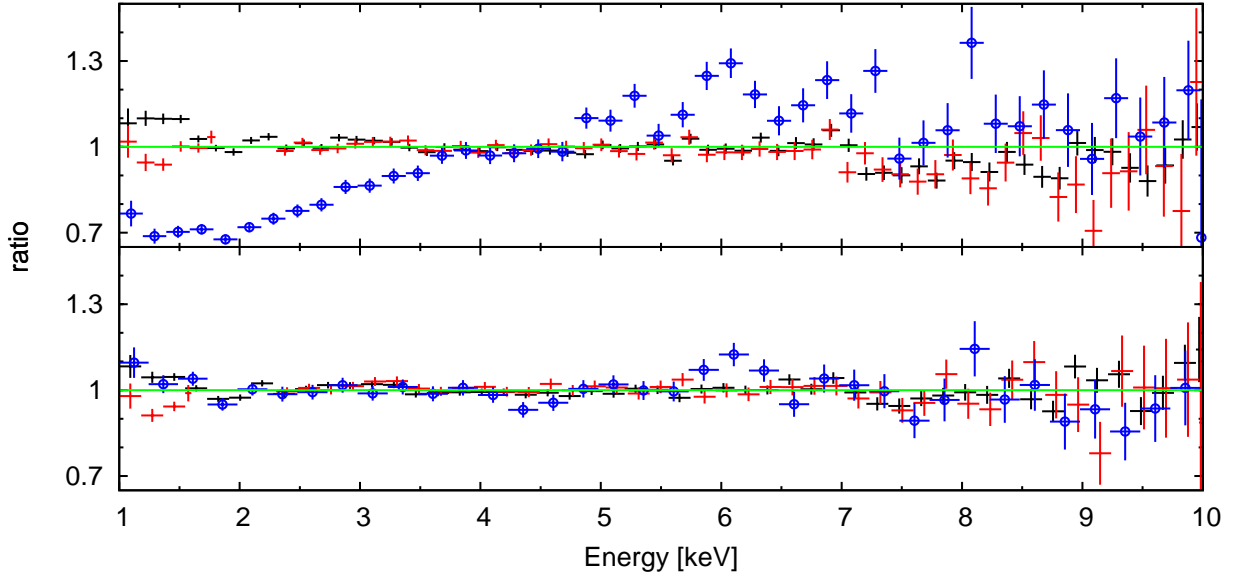}
 \caption{Comparison of {\suzaku} and {\xmm} data residuals.
{\textbf{Up:}} data residuals from the identical underlying model
consisting of an absorbed power law and a Gaussian line.
{\textbf{Bottom:}} data residuals from the final model 
containing an additional absorber
with different covering fraction and slightly
different slope of the power law for the {\xmm}
and the {\suzaku} observation. 
The front-illuminated spectra of {\xis}\,0 and {\xis}\,3 are jointly
shown by black data points, the back-illuminated {\xis}\,1 are red, and
the {\pn} data are blue and emphasised by small circles. 
The data are binned for plotting purposes only.
}
\label{comparison}	
\end{figure*}


\begin{table*}
\caption{Spectral analysis of {\uu} - the model parameters.}
\begin{tabular}{c|c|c|c|c|c|c|c|c|c|c|}
 	\hline \hline \rule{0cm}{0.5cm}
 model component & parameter & \multicolumn{2}{c|}{flat power-law model} &  \multicolumn{2}{c|}{`part. cov.' model}	&
\multicolumn{2}{c|}{complex model `fix'} & \multicolumn{2}{c|}{complex model `free'}	 \\
\hline
\rule[-0.7em]{0pt}{2em} 
	& & {\ekisemem} & {\suzaku} & {\ekisemem} & {\suzaku} & {\ekisemem} & {\suzaku} & {\ekisemem} & {\suzaku} \\
\hline
\rule[-0.7em]{0pt}{2em} 
local absorption & $n_{\rm H, fully} [10^{22}$\,cm$^{-2}]$ & $0.86^{+0.07}_{-0.06}$ & $0.93 \pm 0.02$ & \multicolumn{2}{c|}{$0.92 \pm 0.02$} & \multicolumn{2}{c|}{$0.95 \pm 0.02$}  & \multicolumn{2}{c|}{$0.98 \pm 0.02$} \\
\hline
\rule[-0.7em]{0pt}{2em} 
... & $n_{\rm H, part} [10^{22}$\,cm$^{-2}]$ & \multicolumn{2}{c|}{...} & \multicolumn{2}{c|}{$7.5 ^{+1.3}_{-1.2}$} & \multicolumn{2}{c|}{$9.3 ^{+1.3}_{-1.2}$} & \multicolumn{2}{c|}{$8.2 \pm 1.5$} \\
\hline
\rule[-0.7em]{0pt}{2em} 
... & cov. fraction [\%] & \multicolumn{2}{c|}{...} & $44^{+15}_{-13}$ & $6 \pm 3$ & $50 \pm 3$ & $5 \pm 3$ & $41^{+15}_{-13}$ & $10 \pm 6$ \\
\hline
\rule[-0.7em]{0pt}{2em} 
power law & $\Gamma$ & $1.26^{+0.04}_{-0.02}$ & $1.68^{+0.02}_{-0.01} $ & $1.66 \pm 0.09$ & $1.72 \pm 0.02 $  & \multicolumn{2}{c|}{$1.72^{+0.03}_{-0.02}$}  & $1.68^{+0.08}_{-0.09}$ & $1.83 \pm 0.06$ \\
\hline
\rule[-0.7em]{0pt}{2em} 
reflection & $R$ & \multicolumn{2}{c|}{...} & \multicolumn{2}{c|}{...} & \multicolumn{2}{c|}{$0.48 \pm 0.05$} & \multicolumn{2}{c|}{$2 \pm 1$} \\ 
\hline
\rule[-0.7em]{0pt}{2em} 
... & $E_{\rm fold}$ & \multicolumn{2}{c|}{...} & \multicolumn{2}{c|}{...} & \multicolumn{2}{c|}{$100$ (f)} & \multicolumn{2}{c|}{$200 \pm 100$} \\
\hline
\rule[-0.7em]{0pt}{2em} 
... & $i$ [deg] & \multicolumn{2}{c|}{...} & \multicolumn{2}{c|}{...} &  \multicolumn{2}{c|}{$45$ (f)} & \multicolumn{2}{c|}{$85_{-6}$} \\
\hline
\rule[-0.7em]{0pt}{2em} 
... & $N_{\rm dir} [10^{-3}]$ & $5.7^{+0.3}_{-0.4}$ & $10.9 \pm 0.2$ & $7.1 \pm 0.4$ & $11.0 \pm 0.3$ & $7.6^{+0.7}_{-0.6}$ & $11.2 \pm 0.2$ & $7.5 \pm 0.5$ & $12.2 \pm 0.6$ \\
\hline
\rule[-0.7em]{0pt}{2em} 
... & $N_{\rm abs} [10^{-3}]$ & \multicolumn{2}{c|}{...} & $6 \pm 2$ & $0.7^{+0.4}_{-0.3}$ & $7.6 \pm 0.3$ & $0.6 \pm 0.4$ & $5 \pm 2$ & $1.4 \pm 0.8$ \\
\hline
\rule[-0.7em]{0pt}{2em} 
Gaussian line & $E_{\rm rest}$ [keV] & $6.26^{+0.08}_{-0.07}$ & $6.40 \pm 0.01$ &  $6.29^{+0.06}_{-0.07}$ & $6.40 \pm 0.01$ & \multicolumn{2}{c|}{...} & \multicolumn{2}{c|}{...} \\
\hline
\rule[-0.7em]{0pt}{2em} 
... & $\sigma$ [keV]& $0.23^{+0.09}_{-0.08}$  & $0.08^{+0.02}_{-0.01}$ & $0.21^{+0.06}_{-0.07}$  & $0.08 \pm 0.02$ & \multicolumn{2}{c|}{...} & \multicolumn{2}{c|}{...} \\
\hline
\rule[-0.7em]{0pt}{2em} 
... & $N_{\rm line}$ $[10^{-5}]$ & $10 \pm 3$ & $4.7^{+0.6}_{-0.4}$ & $9 \pm 3 $ & $4.7^{+0.5}_{-0.6}$ & \multicolumn{2}{c|}{...} & \multicolumn{2}{c|}{...} \\
\hline
\rule[-0.7em]{0pt}{2em} 
fit goodness  & $C/\nu$ & \multicolumn{2}{c|}{9543/9110} & \multicolumn{2}{c|}{9484/9108} & \multicolumn{2}{c|}{9477/9114} & \multicolumn{2}{c|}{9460/9111}\\

\end{tabular}

\tablefoot{The common parameters for all the fits are the column density of the absorber
due to interstellar matter in our Galaxy $n_{\rm H, ISM}= 1.06 \times 10^{22}$cm$^{-2}$ \citep{2005A&A...440..775K}, and the cosmological redshift $z=0.012\pm0.002$ \citep{2006A&A...453..839P}. Solar abundances were assumed in the reflection models.
The $N_{\rm dir}$ parameter is a normalisation factor of the direct component 
while $N_{\rm abs}$ corresponds to the component affected by the partially 
covering absorber. 
They belong to the power-law, or the reflection model by {\pexmon}, respectively. 
The sign `(f)' means that the parameter was fixed to that value.}

\label{model} 
\end{table*}

\begin{figure}
 \includegraphics[width=0.49\textwidth]{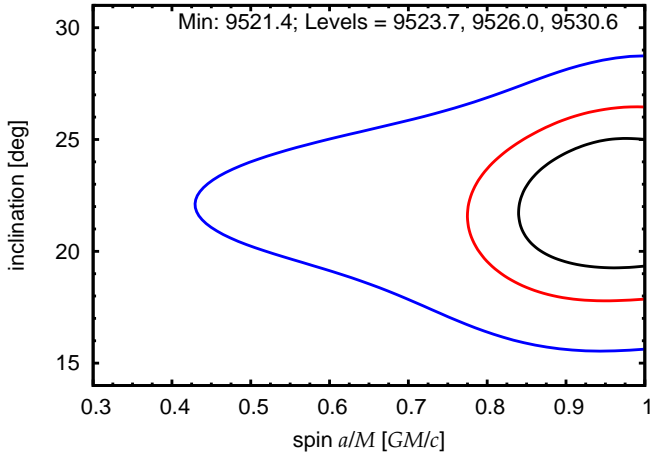}
 \caption{Contour plot of the spin and the inclination angle
measured by the relativistic iron K$\alpha$ line 
in the flat power-law model. The individual lines 
correspond to 68, 90, and 99\% confidence levels.
}
\label{kyrline_contai}	
\end{figure}


\section{Comparison with the {\xmm} observation}
\label{xmm}

\subsection{Evidence of long-term spectral variability}

Although the time-averaged observed flux did not change significantly
between the {\xmm} and {\suzaku} observations
the spectral shape is different. 
The simple power-law model with a Gaussian line applied to both data sets 
with the same parameters as from the {\suzaku} fit
(besides the normalisation factors)
does not provide an acceptable fit to the {\pn} spectrum
with the total $C/\nu = 10476/9112$, 
or {\chired}$=7340/6219 \approx 1.2$, respectively.
Figure~\ref{comparison} (upper panel) clearly
shows that either the photon index of the power law 
or the properties
of the absorber changed between these two observations.

\subsection{Model with a variable power-law slope}

First, we allowed the column density of the local absorber to vary. 
The fit improved
to $C/\nu = 9877/9113$, or {\chired}$=6741/6220 \approx 1.08$, respectively.
The value of the column density 
of the local absorber was found to be $N_{H, {\rm XMM}} = (1.5 \pm 0.1) \times 10^{22}$\,cm$^{-2}$. 
A significantly better fit ($C/\nu = 9554/9111$,
or {\chired}$=6468/6218 \approx 1.04$, respectively) was found
when we also relaxed the power-law photon index. 
New values are $\Gamma_{\rm XMM} = 1.24 \pm 0.04$
and $N_{H, {\rm XMM}} = (0.85 \pm 0.06) \times 10^{22}$\,cm$^{-2}$.
The fit goodness further improved by $\Delta C \approx 11$ 
when, in addition, the iron line parameters were unbound.
The resulting fit parameters are summarised in Table~\ref{model}
in the column labelled ``flat power-law model''.

%


The energy of the iron line is significantly red-shifted
towards its rest energy and broadened, consistently with the findings
by \citet{2006A&A...453..839P}.
We replaced the Gaussian line by the relativistic iron line model
{\textsc{kyrline}} \citep{2004ApJS..153..205D}.
The resulting fit improved by $\Delta C \approx 22$
with respect to the model with a broad Gaussian line,
suggesting significant asymmetry of the line profile.
The angular momentum of the black hole is preferred to be high,
i.e. the disc extends very close to the black hole.
The inclination angle was found to be $i = (22 \pm 7)^\circ$.
We assumed the standard radial emissivity profile 
(i.e. reflection flux $\propto r^{-3}$)
and isotropic angular emissivity.
Figure~\ref{kyrline_contai} shows the confidence levels
of the spin and the inclination angle.

Although this model can satisfactorily explain
the {\xmm} spectrum, the narrow line from the {\suzaku}
spectrum suggests that the reflection originates
much further from the central black hole.
A major part of the iron-line profile can be indeed
explained by the same narrow line as seen in the {\suzaku}
spectrum (see Sect.~\ref{xmm_iron} for more discussion).
Therefore, we proposed an alternative model explaining
the spectral change between the {\xmm} and {\suzaku} observation.

\subsection{Model with a complex absorber's variability}

\citet{2006A&A...453..839P} suggested
that a partially covering absorption might be responsible 
for measuring the flat spectral shape in the {\xmm} observation.
Employing the model with more complex absorption did not improve the statistical
goodness of the fit, but yielded a photon index more similar
to the values measured in Seyfert galaxies \citep[see e.g.][]{2009A&A...495..421B}.
%
Therefore, we also included the partially covering absorption in our model,
first only to the {\xmm} data.
We bound the iron-line
parameters between the {\suzaku} and {\xmm} spectra
for the case that the spectral curvature at around 5-6\,keV 
could be explained by partially absorbing clouds, similarly as 
in spectroscopy studies of different sources
by \citet{2008A&A...483..437M} or \citet{2011ApJ...733...48T}.

The photon index of the {\xmm} spectrum, $\Gamma = 1.6 \pm 0.1$, 
is indeed found to be more consistent with the {\suzaku} observation.
The goodness of the fit is also better than in the previous case 
with $C/\nu = 9499/9109$, or {\chired}$=6408/6216 \approx 1.03$, respectively.
The column density of the fully covering absorption
was found to be $N_{H, {\rm XMM, fully}} = 0.86 \pm 0.11 \times 10^{22}$\,cm$^{-2}$.
The fraction of the partially absorbed power law is $44 \pm 12\,\%$
with the column density $N_{H, {\rm XMM, part}} = 7.1^{+2.0}_{-1.8} \times 10^{22}$\,cm$^{-2}$.

As a next step, we included the partially covering absorption 
also for the {\suzaku} data. 
We simply assumed that only the fraction of the partially covering 
absorption changed between the times of the two observations 
(allowing the column densities to vary
did not improve the fit significantly).
The column density of the fully covering absorption
was found to be $N_{H, {\rm fully}} = 0.92 \pm 0.02 \times 10^{22}$\,cm$^{-2}$,
the column density of the partial covering 
$N_{H, {\rm part}} = 7.5^{+1.3}_{-1.2} \times 10^{22}$\,cm$^{-2}$,
and the fraction of the partially absorbed power law 
$44^{+15}_{-13}\,\%$ for the {\xmm} and $5 \pm 3\,\%$ 
for the {\suzaku} spectrum.
The fit goodness is $C/\nu = 9490/9110$,
or {\chired}$=6347/6217 \approx 1.02$, respectively.

\begin{figure}
 \includegraphics[width=0.49\textwidth]{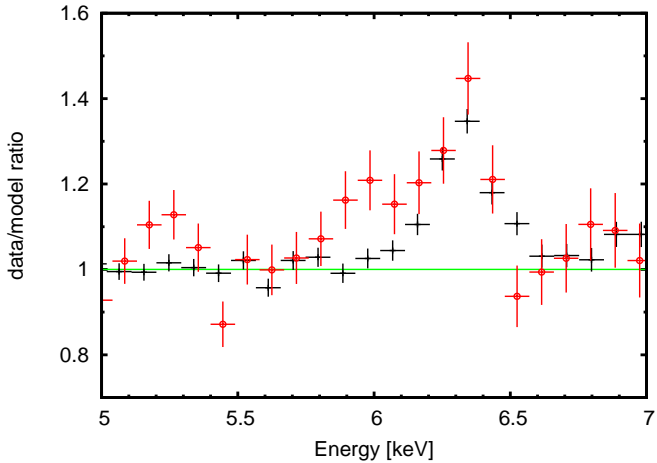}
 \caption{Spectral residuals in the iron line energy range.
The residuals are calculated against
model consisting of power law with the partially covering absorber.
Black data points correspond to the spectra of 
front-illuminated {\suzaku}/{\xis} detectors.
The red ones (with small circles) correspond to {\pn} spectrum.
The data are binned for plotting purposes only.}
\label{iron_line}	
\end{figure}

\citet{2006A&A...453..839P} rejected the model with the partially
covering absorber (their ``model E''), 
because it did not explain all the residuals at the iron-line energy band.
Indeed, when we unbound the iron-line parameters,
we obtained similar results as with the flat-powerlaw model.
The energy of the line is found red-shifted,
$E=6.3\pm 0.1$\,keV, and broadened, $\sigma=0.2\pm 0.1$\,keV.
The equivalent width of the line is
$EW_{\rm SUZ} = 96^{+10}_{-12}$\,eV with {\suzaku},
and $EW_{\rm XMM} = 160^{+60}_{-50}$\,eV with {\xmm}.
The difference consists in the width of the line.  
A more detailed look on the iron-line residuals is shown in Fig.~\ref{iron_line}.
The partially-covering absorber does not explain
why the iron line is broader and more redshifted
in the {\xmm} spectrum.
However, we consider this model further, as it is more consistent
with the {\suzaku} measurements
than the flat power-law model.

\subsection{More complex reflection models}

The iron line detected in the {\xis} spectra implies 
the presence of the cold reflection component in the spectrum.
We used further {\pexmon} model \citep{2007MNRAS.382..194N},
which combines the reflection continuum by {\pexrav} model \citep{1995MNRAS.273..837M}
with the appropriate strength of the fluorescent lines of iron and nickel.
Because the iron line detected by {\xis} detectors
is narrow, the reflection must occur sufficiently far
from the black hole and is rather unlikely to be variable
between the two observations. 
Therefore, we simply assumed the same reflection
model for both {\suzaku} and archival {\xmm} data.

Neither the inclination angle nor the origin of the reflection is known
in this galaxy. \citet{2006A&A...453..839P} determined the type of 
Seyfert 1.5 galaxy from the optical measurements. In terms of the AGN
unification scenario by \citet{1993ARA&A..31..473A}, this would correspond
to an intermediate value of the inclination, i.e. $i \approx 45$ degrees.
Therefore we considered two cases: first one with keeping the
values of the inclination and the folding energy fixed in the fit
(called as ``complex fix'' in Table~\ref{model}) 
and second one with these parameters free (``complex free'' in Table~\ref{model},
or ``final'' model hereafter).

\begin{figure}
 \includegraphics[width=0.49\textwidth]{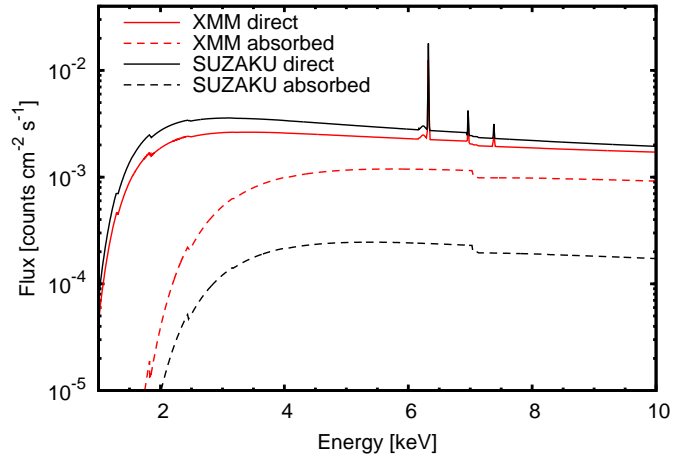}
 \caption{Final model consisting of a partially covering absorber
and cold reflection. The model is shown by red or black related to the {\xmm} 
or the {\suzaku} observation, respectively. The components affected by the
partially covering absorber are shown by dashed lines while the direct components
by solid lines.}
\label{final_emodel}	
\end{figure}

The fit goodness of the final model is characterised by $C/\nu = 9460/9111$,
or {\chired}$=6316/6218 \approx 1.02$, respectively.
%
The final model is plotted in Figure~\ref{final_emodel}
in the $1-10$\,keV energy range. It is demonstrated there
how the fraction affected by the partially covering absorber
decreased between the two observations.
The iron and nickel fluorescent lines as well as the iron edge
are the properties of non-relativistic cold reflection.

Although the spectrum is described rather satisfactorily
by the final model, 
the residuals at the energy range around 6\,keV still 
persist in the {\xmm} spectrum.
The data/model ratio is shown in the bottom panel 
of Fig.~\ref{comparison}.
In order to demonstrate the significance of the remaining residuals,
we added a Gaussian line to the final fit.
The fit improves by $\Delta C = 22$
(or $\Delta \chi = 23$, respectively). 
This is a statistically significant result
as the random improvement should be of order of $\Delta C \approx 6$
for the three extra parameters added to the model.
The energy of the Gaussian line was found to be
$E= 6.15^{+0.12}_{-0.11}$\,keV, the width $\sigma=0.21^{+0.12}_{-0.11}$\,keV, 
and normalisation factor $N_{\rm Gauss} = 5.8^{+1.9}_{-2.1} \times 10^{-5}$.
The equivalent width of this feature is $90^{+100}_{-40}$\,eV.
This additional line can be attributed to a redshifted iron line 
emitted from inner parts of the accretion disc
-- see Sect.~\ref{xmm_iron} for more discussion
on this spectral feature.


\begin{table*}
\caption{Flux of {\uu} measured by different instruments
and predicted values by various models.}
\begin{center}
\begin{tabular}{c|c|c|c|c|c}
 	\hline \hline \rule{0cm}{0.5cm}
\rule[-0.7em]{0pt}{2em} instrument (-- model) & \multicolumn{5}{c}{flux [$10^{-11}$\,erg\,cm$^{-2}$\,s$^{-1}$]}\\
\hline
\rule[-0.7em]{0pt}{2em} &	$2-10$\,keV (observed) &	$2-10$\,keV (intrinsic)	&	$20-40$\,keV 	&	$40-100$\,keV 	&	$20-100$\,keV 	 \\
\hline
\rule[-0.7em]{0pt}{2em} {\it XMM} - flat $\Gamma$ & $4.16 \pm 0.04$ & $4.71 \pm 0.04$ & $7.7 \pm 0.1$ (p) & $19.0 \pm 0.2$ (p) & $26.7 \pm 0.3$ (p)\\
\hline
\rule[-0.7em]{0pt}{2em} {\it XMM} - part. cov. & $4.04 \pm 0.04$ & $5.56 \pm 0.04$ & $4.3 \pm 0.1$ (p)	& $7.5 \pm 0.1$ (p) & $11.9 \pm 0.2$ (p)\\
\hline
\rule[-0.7em]{0pt}{2em} {\it SUZAKU} - part. cov. & $3.83 \pm 0.01$ & $4.58 \pm 0.02$ & $3.29 \pm 0.02 $	& $5.49 \pm 0.03 $ (p) & $8.75 \pm 0.05$ (p)\\
\hline
\rule[-0.7em]{0pt}{2em} {\it SUZAKU} - final & $3.83 \pm 0.01$ & $4.67 \pm 0.02$ & $3.35 \pm 0.02 $	& $4.09 \pm 0.03 $ (p) & $7.44 \pm 0.05$ (p)\\
\hline
\rule[-0.7em]{0pt}{2em} {\integral} & ... & ... & $3.2 \pm 0.1 $\tablefootmark{a}	& $4.1 \pm 0.2 $\tablefootmark{a} & $6.6$\tablefootmark{b}\\
\hline
\rule[-0.7em]{0pt}{2em} {\swift} & ... & ... & $2.7 \pm 0.1$	& $4.0 \pm 0.2 $ & $6.7 \pm 0.3$\\
\hline
\end{tabular}
\end{center}

\tablefoot{
A sign (p) means that the value was predicted
by the model from lower X-ray energies.
\tablefoottext{a}{\citet{2010ApJS..186....1B}},
\tablefoottext{b}{\citet{2008A&A...483..151P}}.
}
\label{high_flux} 
\end{table*}

\section{Discussion}
\label{discussion}


\subsection{Presence of a partially covering absorber}

The {\suzaku} observation of {\uu} reveals that the spectrum
can be characterised by a photon index of the power law $\Gamma \approx 1.7$,
which is a typical value for Seyfert galaxies \citep{2009A&A...495..421B}.
The different slope of the {\xmm} spectrum can be explained by 
a very low photon index ($\Gamma \lesssim 1.3$)
or by the presence of a partially covering absorber. 
A statistically better fit was obtained with the model
with the partially covering absorber.
Moreover, this model yields flux at high energies more consistent
with the measured values by the {\integral} and {\swif} satellite,
as demonstrated in Table~\ref{high_flux}.

%
High-energy flux predicted by different models 
(flat power law, model with the partially covering absorber,
final model)
is compared with the values measured 
by the {\integral} and {\swif} satellite\footnote{A power-law model was used to measure {\swift} 
flux from the 58-month hard X-ray survey data. http://heasarc.gsfc.nasa.gov/docs/swift/results/bs58mon/}.
Better agreement is obtained
for the scenario with the partially covering absorber
(best for the final model).
However, this cannot be regarded as a final proof owing to
a possible long-term variability between the non-simultaneous
observations. Moreover, a low high-energy cut-off in the flat
power law would significantly decrease the predicted
hard X-ray flux by this model.

An additional fully-covering absorber with $N_{H} \approx 10^{22}$\,cm$^{-2}$
along our line of sight to the nucleus of {\uu} was required by the data.
It may be identified with dust lanes or star-forming regions in the innermost 
host galaxy disc and not associated with the AGN torus \citep{2000A&A...355L..31M}.
A very large contamination from the host galaxy has also been
claimed by \citet{2010MNRAS.402.1081V} in order to explain
the infrared spectral energy distribution of {\uu}.
This scenario is consistent with the intermediate type 
classification \citep[see e.g.][]{1995ApJ...454...95M},
which was originally proposed for {\uu} by \citet{2006A&A...453..839P}.


\subsection{The iron line in the {\xmm} observation}
\label{xmm_iron}

Model consisting of a power-law component modified by the fully
and partially covering absorber and the narrow iron fluorescent line
represents a good simultaneous fit to both, the {\suzaku} and the archival {\xmm}
spectrum, apart from the emission residuals
at the moderately redshifted iron line energy band 
that are entirely related to the {\xmm} spectrum.
Figure~\ref{iron_line} shows this energy band
for the {\suzaku}-{\xis} and {\xmm} spectrum in detail. 
\citet{2006A&A...453..839P} concluded from the {\xmm} spectrum
that the total iron line profile is most likely formed 
in the innermost accretion disc.
However, the line profile measured by {\suzaku} is much narrower. 
Therefore, we interpreted the {\xmm} iron line profile
as the narrow component (the same as in the {\suzaku} spectrum) 
plus an additional enhanced emission 
from the innermost accretion disc.





Such an enhanced emission can be explained, e.g., by a temporary irradiation of a part 
of the accretion disc by a flare,
for instance due to a magnetic reconnection in the corona 
\citep{2004A&A...420....1C, 2008ApJ...682..608U}.
With the energy $E \approx 6.1$\,keV,
it is significantly redshifted when associated to the iron fluorescent line.
Therefore, the probable origin is from the innermost accretion disc
where the radiation is affected by strong gravitational redshift.
A transient emission feature at a similar energy was reported
for another Seyfert galaxy NGC\,3516 
\citep{2002ApJ...574L.123T, 2004A&A...422...65B}.
Much longer exposure time allowed to trace the temporal evolution of the emission
and reveal possible periodicity due to an orbiting spot in the accretion disc
\citep{2004MNRAS.355.1073I}.

We performed several tests using {\ky} model \citep{2004ApJS..153..205D}
including also non-axisymmetric models (due to the presence
of a partially covering absorber).
We tried to split the {\xmm} observation into several parts
and look for the change of residuals. We also generated a light curve
related only to the redshifted component of the iron line
and compared it with the total light curve.
However, the insufficient data quality of the 
short {\xmm} observation does not allow
us to uniquely constrain neither 
the geometrical structure of the emission region
nor its temporal behaviour. 
It was not possible to distinguish between
a ``ring-shaped'' static emitting region and an orbiting spot.
The evidence of such spectral features, however,
satisfy the studies of a locally enhanced emission from parts of
the accretion disc \citep{2004MNRAS.350..745D, 2007A&A...475..155G},
and will be potentially resolved with future higher resolution
X-ray instruments \citep[see e.g.][]{2011MNRAS.418..276S}.

The absence of any relativistic features in the {\suzaku} spectrum
is rather puzzling. Although the {\uu} observation revealed
a complex absorption the {\suzaku} observation provided us the clearest
view on the nucleus. 
A possible explanation for non-detection of the redshifted
iron-line emission could be due to a truncation of the disc
at a further radius.
A truncated accretion disc was reported for 
several other sources 
\citep[see e.g.][]{2005A&A...435..857M, 2009ApJ...705..496M, 2010A&A...512A..62S}.
The physical reason for such a truncation might be due to a very low accretion
rate when the accretion flow would be advection dominated
and would not form a thin accretion disc \citep{1997ApJ...489..865E}, 
or due to an over-ionisation of the innermost part of the accretion disc
by strong irradiation \citep{2000ApJ...536..213D,2011MNRAS.416..629B}.

The change of the iron line from a relativistically broadened
profile to a narrow one as a likely result of the state transition of the source,
was reported for a quasar Q0056-363 \citep{2005A&A...435..857M}. 
Different photon indices were
measured between the two observations but on contrary to our case
the flatter index was related to the narrow line detection.
An explanation in terms of the disc 
truncation is therefore unlikely in the case of {\uu}
since it does not suite well in a truncated disc scenario in the hard states
of Galactic black hole binaries \citep{2001MNRAS.323L..37L, 2004MNRAS.355.1105F}. 
Moreover, the measured flux during the {\suzaku} observation is very close
to that measured by {\xmm}.
The accretion rate derived from the infrared
and X-ray observations is $\lambda \approx 0.03 \lambda_{\rm Edd}$ \citep{2010MNRAS.402.1081V},
which is somewhat typical value for other Seyfert galaxies as well.

Therefore, the ionisation may represent a more plausible explanation.
The illuminating flux is a strongly radially dependent function,
especially if the corona is very compact \citep{1993MNRAS.262..179M, 2008ApJ...675.1048R, 2012arXiv1208.0360S}.
The innermost disc can be over-ionised owing to the combination of sufficiently high
irradiation flux and relatively low density of the accreting material.
The ionisation state could, however, locally decrease
if the density is locally increased (e.g. due to a shock wave), 
or if some clouds partly obscure the illuminating source.
The latter one could happen during the {\xmm} observation.
As a result, some additional fluorescent emission
was detected from the innermost part of the accretion disc.


An alternative explanation for the missing imprints of the innermost region
could be due to a distant corona.
Speculatively, if the scattering medium is associated with the
partially covering absorber this would also explain the change
between the {\xmm} and the {\suzaku} observation.
The decreased covering fraction would be due to 
an out-flowing corona and thus, less emission
from the innermost part of the disc would be detected
in the {\suzaku} observation.
Smaller intrinsic X-ray flux (see Table~\ref{high_flux})
during the {\suzaku} observation would match well this scenario.



\begin{figure}
 \includegraphics[width=0.49\textwidth]{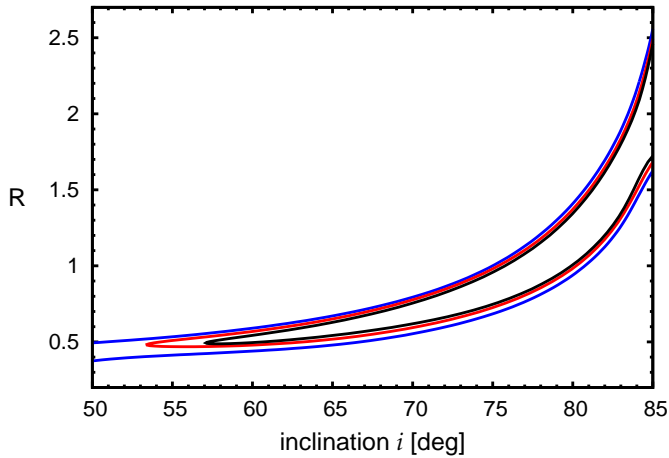}
\caption{Contour plot between the inclination angle and reflection parameter
of the final model. The individual contours correspond to 68, 90, and 99\%
confidence levels.}
\label{contir}	
\end{figure}


\subsection{Reflection models and the angle of inclination}
\label{discussion_ref}

The reflection model {\pexmon} was used for a consistent description
of the iron K$\alpha$, the iron edge and the Compton hump.
The reflection fraction $R$, the folding
energy $E_{\rm fold}$ and also the inclination angle $i$  
are, however, not constrained very well and suffer from a mutual degeneration
of the parameters. 
Moreover, the link between the Compton hump
and fluorescent lines might not be necessarily fulfilled.
Especially in the case of reflection
from the torus a different geometry should be taken
into account \citep{2009MNRAS.397.1549M, 2012MNRAS.423.3360Y}. 
We also performed several test fits with the {\pexrav} model 
and a Gaussian line separately
and we get almost identical results like with {\pexmon},
i.e. favouring a high inclination angle 
measured relatively to the normal of the orbital plane.

However, it is necessary to stress that 
the inclination angle intervenes only as a projection effect in the 
{\pexrav} or {\pexmon} model. The shape of the emission 
is independent of the emission angle 
and the total spectrum is averaged over all angles.
As a result, both $R$ and $i$ affect
only the normalisation and are therefore degenerated.
This is apparent
in Figure~\ref{contir}
where the confidence levels of the reflection parameter $R$
and the inclination angle $i$ of the final model are shown.
Illumination of the accretion disc by an isotropic source on top of it
corresponds to the value $R=1$.
The larger values of $R$ would suggest that the more observed radiation comes 
from reflection than from the direct component.

The resulting high inclination angle of the final model,
which is also suggested for the case $R=1$, is likely a consequence
of the low contrast between the Compton scattering hump
and the primary continuum, the former being determined
by the combination of strong Compton scattering
and low photoelectric absorption cross-section.
The lower inclination would imply the Compton hump
to be more prominent given the observed equivalent 
width of the iron line. 

This would would imply two possible scenarios. 
First, the system could be oriented almost
face-on but we see the reflection on the 
distant broad line region or the torus, which gives a complementary
reflection angle.
This is supported by the detection of broad lines in the optical spectrum
\citep{2006A&A...453..839P}.
It is also consistent with low inclination angle of the innermost
accretion disc measured by the relativistic iron line model
in the flat power-law scenario
for the {\xmm} spectrum (see Fig.~\ref{kyrline_contai}).
The second scenario assumes that
the reflection occurs at the more distant parts of an accretion disc 
which is oriented almost edge-on relatively to the observer.

The edge-on disc hypothesis is not in a strong contradiction 
with the optical measurements.
The standard unification scenario
has been questioned in several sources, 
for a review see \citet{2012AdAst2012E..17B} and references therein.
The clumpiness of the absorber is supposed to be another important parameter
in the proper classification and the type of AGN could not be entirely
determined by the inclination angle but also by a probability
of the absorption clouds intercepting the line of sight \citep{2002ApJ...570L...9N, 2012ApJ...747L..33E}. This is supported by a large variability
of the absorber observed in several sources \citep[see e.g.][]{2005ApJ...623L..93R}.
{\uu} has also revealed partially covering absorber
with a significant change between the archival {\xmm}
and recent {\suzaku} observation, and can be thus another
example advocating for an updated unification model.

Moreover, the inclination angle of the accretion disc
and the host galaxy might be misaligned \citep[see e.g.][]{2011A&A...531A.131G}. 
An evident misalignment of the circum-nuclear matter with
the radio jet, which is naturally believed to lie in the black hole
rotation axis, was reported by \citet{2009MNRAS.394.1325R}.
\citet{2011MNRAS.411.2223R} analysed the distribution
of the [\ion{O}{iii}] emission line equivalent width in order to
constrain the inclination in a large sample of AGNs, 
and reported several sources with edge-on discs but 
not heavily absorbed by a circum-nuclear matter.
{\uu} could belong to this kind of sources as well.

\section{Conclusions}
\label{conclusions}

The X-ray continuum spectrum of the Seyfert galaxy 4U 1344-60 
is dominated by a power law with a standard value of the photon index 
$\Gamma \approx 1.7 \pm 0.1$.
A narrow fluorescent line with equivalent width $\approx 100$\,eV
is an apparent feature in the {\xis} spectra.
The width of the iron line is $\sigma=(0.08 \pm 0.02)$ keV,
i.e. $\approx 9000$ km/s of FWHM,
suggesting its origin from a distant matter not affected by relativistic smearing.
The outermost parts of an accretion disc, broad line region
clouds or torus are possible candidates for the reflector. 
Its profile is, however, significantly narrower than in the archival
{\xmm} observation.
The detected red-shifted iron line emission 
during the {\xmm} observation was
likely a temporary feature
either due to locally enhanced emission
or decreased ionisation 
in the innermost accretion flow.

The spectral shape has also changed between the {\xmm} and {\suzaku} observation.
We interpret this change as due to the presence of a partially covering absorber. 
The absorbed fraction of the primary radiation 
has decreased from around $45\%$ to less than $10\%$.
The resulting model is then characterised by the same power law slope.
It is also consistent with the flux measured (non-simultaneously)
by the {\integral} and {\swif} satellite at high X-ray energies.
The spectral variability makes {\uu} 
an interesting target for further examination
within a monitoring programme by an X-ray satellite.

\begin{acknowledgements} 

The work was supported from
the grant ME09036 in the Czech Republic (VK).
JS acknowledges discussion with Andy Pollock about
the statistical processing of the data
and with other colleagues attending Scientific Mondays at ESAC.

\end{acknowledgements}


\bibliographystyle{aa} 
\bibliography{svoboda} 


\vspace{-0.1cm}
\section{Appendix A}

\subsection{Timing and spectral properties of Centaurus B from ASCA to Suzaku}

\begin{figure*}[tb]
\begin{center}
\includegraphics[width=0.47\textwidth]{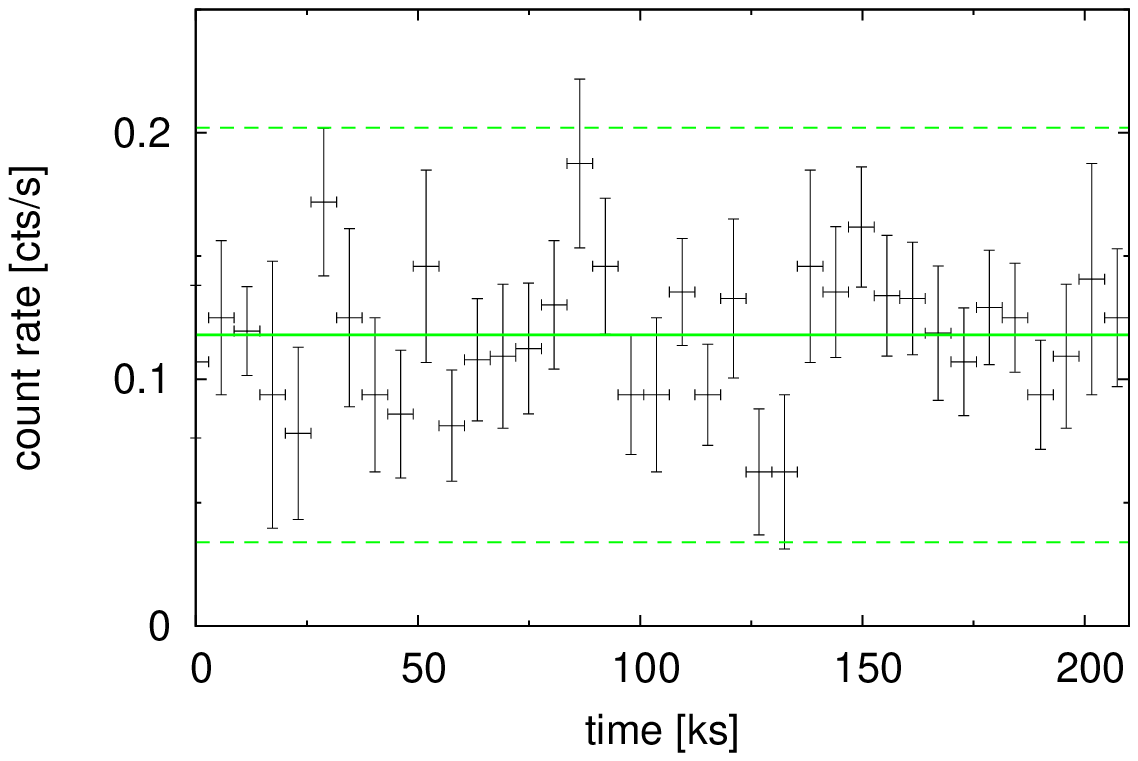}
\includegraphics[width=0.48\textwidth]{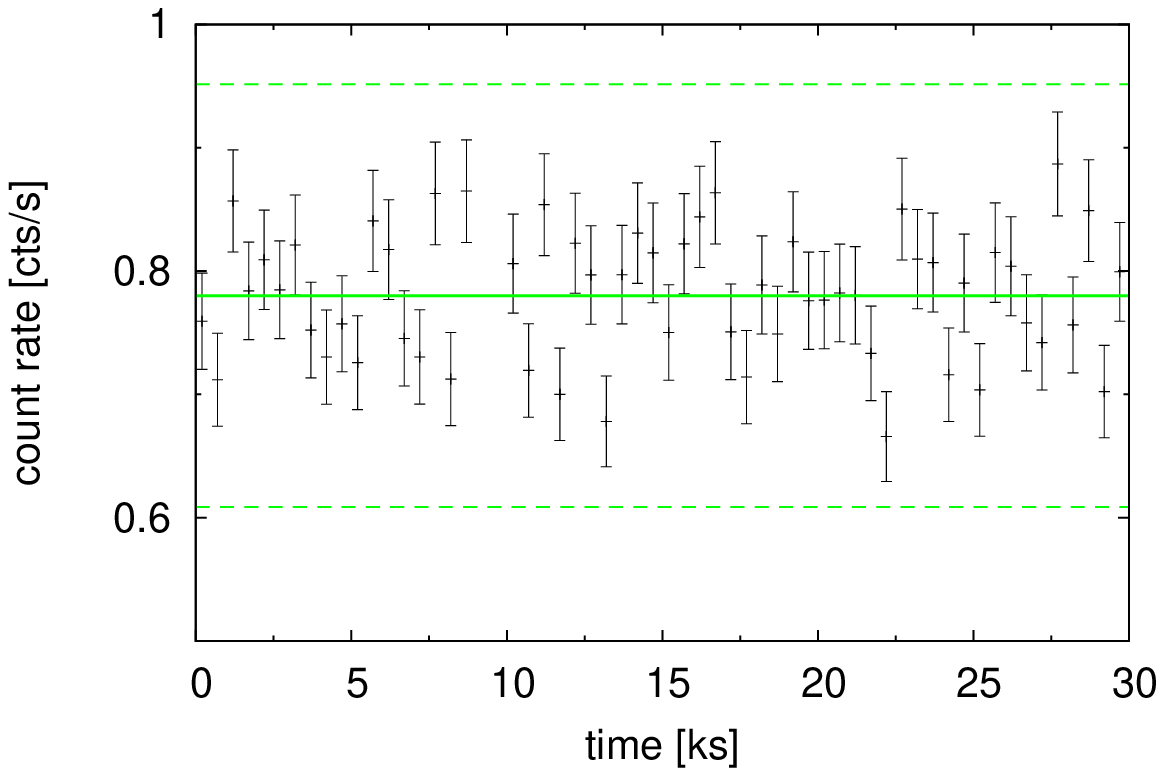}
\vspace{-0.6cm}
\end{center}
\caption{Light curve of Centaurus\,B. {\textbf{Left}}: {\asca} (200\,ks).  
{\textbf{Right}}: {\xmm} (30\,ks). 
Horizontal solid lines represent the average flux.
Dashed lines show the $3\sigma$ levels
where $\sigma$ is the expected standard deviation 
due to Poisson noise.
}
\label{lc_ascaxmm}
\end{figure*}

{\cenb} is one of the nearest and brightest radio galaxies,
at cosmological redshift $z=0.01215$ and flux density
$\approx 250\,Jy$ at 408\,MHz \citep{2001MNRAS.325..817J}.
It is located only at $13'$ from 4U\,1344-60 on the sky. 
The proximity of the two sources did not allow us to set an observing
configuration in which we can exclude Centaurus\,B from the HXD
detectors. Therefore, we required an accompanying $10$\,ks observation
targeted directly on Cen B to estimate its X-ray flux in {\hxd} energy range. 
Here, we present our spectral results from this short
{\suzaku} observation using the {\xis} detectors 
together with the analysis of its long-term variability.
The data reduction procedure is the same as described
in Section~\ref{data}.

Historically, the flux of {\cenb} was found to be only a factor of a few 
smaller than 4U\,1344-60.
The observed flux in the 2-10~keV was
$7.3 \pm 0.2 \times10^{-12}$~erg cm$^{-2}$ s$^{-1}$ in the archival {\asca} observation
\citep{1998ApJ...499..713T},
and $5.6 \pm 0.2 \times10^{-12}$~erg cm$^{-2}$ s$^{-1}$ in the {\xmm} observation
\citep[see also][]{2005xrrc.procE7.24T}, 
i.e. around $6-7$ times fainter than 4U\,1344-60.
The Integral satellite measured hard X-ray fluxes  of
$f_{20-40\,{\rm keV}}=(0.76 \pm 0.08) \times10^{-11}$~erg\,cm$^{-2}$\,s$^{-1}$
and $f_{40-100\,{\rm keV}}=(1.1 \pm 0.2) \times10^{-11}$~erg\,cm$^{-2}$\,s$^{-1}$ 
\citep{2010ApJS..186....1B}, around $4$ times less than the 4U 1344-60 fluxes.  

Figure~\ref{lc_ascaxmm} shows X-ray light curves 
from the archival {\asca} and {\xmm} observations. 
The light curves look stationary with no trend in the flux.
We fitted the light curves by polynomial fits, and we found that 
a constant function represents the best fit.
Most of the apparent variability in the light curves 
is due to Poisson noise whose level is shown by the dashed lines in the plots.
These results clearly suggest very low variability
and, in particular, no trend in the average flux
of Cen\,B  on a $\approx 100$\,ks time-scale.

The 2-10~keV flux measured with the {\suzaku}/{\xis} detectors is
$(3.9 \pm 0.2) \times10^{-12}$~erg cm$^{-2}$ s$^{-1}$,
which follows a continuous decrease of the flux since
the ASCA observation (see Table~\ref{time_evolution_flux}).
The extrapolated 15-60\,keV flux is then approximately 
$8\times10^{-12}$~erg\,cm$^{-2}$\,s$^{-1}$,
which is one order of magnitude less than 4U\,1344-60.
Thus, in the hard X-ray regime, the flux of 4U\,1344-60 clearly dominates.

\begin{figure}[tb]
 \includegraphics[width=0.49\textwidth]{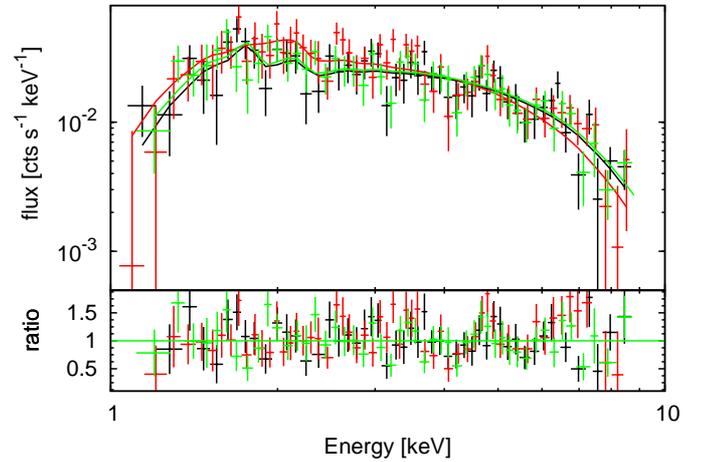}
 \caption{X-ray spectrum of Centaurus B observed with {\suzaku} {\xis}\,0 (black), {\xis}\,1 (red), {\xis}\,3 (green).}
\label{cenb_xis}	
\end{figure}

Although the X-ray brightness of Cen\,B has decreased by a factor
of two from the archival {\asca} observation in 1995 
to the recent {\suzaku} observation in 2011 
the spectral shape has not changed (see Table~\ref{time_evolution_flux})
and can be characterised
by a simple power law with the photon index $\approx 1.6$
and local absorption with the column density
$N_{H} \approx 0.7 \times 10^{22}$cm$^{-2}$.
The {\suzaku}/{\xis} spectrum of Centaurus B 
is shown in Fig.~\ref{cenb_xis}.
The underlying model is {\tbabs}*{\ztbabs}*{\zpowerlaw} in the XSPEC notation
with the best-fit values from the Table~\ref{time_evolution_flux}.
The statistical goodness of the fit can be characterised
by $C/\nu = 177/162$ (or $\chi^2/\nu = 169/162 \approx 1.04$).
%
Some residuals from the model are apparent in the spectrum 
but they are likely due to a less than perfect subtraction 
of the background, considering the relatively low flux of CenB
(the fraction of the background flux is around 30\%).





\begin{table}[tb]
\caption{Evolution of the X-ray flux and spectral slope of Centaurus B.}
\begin{tabular}{c|c|c|c}
 	\hline \hline \rule{0cm}{0.5cm}
  	&	{\asca}	&	{\xmm}	&	SUZAKU	 \\
\hline
\rule[-0.1em]{0pt}{1em} $2-10$\,keV flux & & & \\ 
\rule[-0.1em]{0pt}{0em} [$10^{-12}$\,erg\,cm$^{-2}$\,s$^{-1}$] 
& $7.3 \pm 0.2$\tablefootmark{a}	& $5.6 \pm 0.2$ & $3.9 \pm 0.2$\\
\hline
\rule[-0.7em]{0pt}{2em} photon index 
& $1.64 \pm 0.07$\tablefootmark{a}	& $1.6 \pm 0.1$ & $1.6 \pm 0.2$\\
\hline
\rule[-0.7em]{0pt}{2em} column density\tablefootmark{b} 
& $1.76 \pm 0.11$\tablefootmark{a}	& $1.7 \pm 0.1$ & $1.7 \pm 0.3$\\
\end{tabular}

\tablefoottext{a}{Tashiro et al., 98}\\
\tablefoottext{b}{including Galactic absorption ($N_{H} = 1.06 \times 10^{22}$cm$^{-2}$)}\\


\label{time_evolution_flux} 
\end{table}




\subsection{Estimate of the contamination of the {\uu} {\hxd} spectrum by {\cenb}}
\label{hxd}


Although the high energy flux of {\cenb} is one order
of magnitude lower, we considered it for the proper cross-normalisation
between {\xis} and {\hxd} instruments.
We performed a fit of {\hxd} spectrum of {\uu}
with a single and double power-law model. 
The second power-law component did not improve 
the fit at all. However, a slightly flatter photon index 
was preferred using the single power-law model.
This would be an expectable influence of the contamination
by a harder {\cenb} spectrum with the photon index
 $\Gamma \approx 1.6 \pm 0.2$.
Although the determined error of the photon index 
from the short Suzaku observation is rather high,
the value of $1.6$ has been consistently measured 
since the earliest ASCA observation (see Table~\ref{time_evolution_flux}).

We used the best fit
values obtained from the analysis of the {\xis} spectra of {\cenb}
multiplied by a constant between 0 and 1. 
We fitted the {\xis} and {\hxd} spectra simultaneously 
with the final C-value $7847$ with $7505$ degrees of freedom,
or {\chired}$=5839/5645$, respectively.
This means only a marginal improvement with $\Delta C \approx 3$
(or $\Delta \chi \approx 4$, respectively)
compared to the model without any contamination taken into account.
The best-fit value for the constant is $c= 0.1^{+0.4}_{-0.1}$.
Given the fact that the flux of {\cenb} is one order of magnitude
weaker than for {\uu} the contamination is 5\% at maximum.

\end{document}